\documentclass[10pt,letterpaper,twocolumn]{article} 

\usepackage{ol2}
\usepackage[draft]{hyperref}
\usepackage{amsmath}

\begin{document}

\twocolumn[ 

\title{Correlation-stability approach to elasticity mapping   in OCT: comparison with displacement-based mapping and \textit{in vivo} demonstrations}


\author{V.Yu. Zaitsev,$^*$ L.A. Matveev, G.V. Gelikonov, A.L. Matveyev, and V.M. Gelikonov}

\address{
Institute of Applied Physics RAS, Uljanova St. 46, Nizhny Novgorod, 603950, Russia\\
$^*$Corresponding author: vyuzai@hydro.appl-sci.nnov.ru
}

\begin{abstract}A variant of compression optical coherence elastography for mapping relative tissue stiffness is reported.  Unlike conventionally discussed displacement-based (DB) elastorgaphy, in which the decrease in the cross-correlation is a negative factor causing errors in mapping displacement and strain fields, we propose to intentionally use the difference in the correlation stability (CS) for deformed tissue regions with different stiffness.  We compare the parameter ranges (in terms of noise-to-signal ratio and strain) in which the conventional DB- and CS-approaches are operable. It is shown that the CS approach has such advantages as significantly wider  operability region in terms of strain and is more tolerant to noises. This is favorable for freehand implementation of this approach. Examples of simulated and real CS-based elastographic OCT images are given. \end{abstract}

\ocis{(110.4500) Optical coherence tomography; (170.6935) Tissue characterization.}
 ] 

\noindent The problem of elasticity imaging in optical coherence tomography (OCT) has been attracting much attention since the end of 90ies \cite{Schmitt1998, Rogowska2004, Kirkpatrick2006, Bossy2007,Adie2010, Kennedy2011,  Sun2011, Wang2006, Kennedy2011b}. 
However, the elastography regime is not yet realized in commercial OCT scanners, whereas in medical ultrasonics combined conventional scans and shear-elasticity imaging are already implemented in several platforms (e.g., Siemens, Hitachi, Ultrasonix  \cite{Parker2011}). Although some ultrasonic systems use excitation of shear waves  and measuring their velocities (e.g., Fibroscan and Supersonic scanners)  \cite{Parker2011}, mostly the elasticity imaging is based on special processing of conventional ultrasonic images. 
Since the first works on OCT-elastography \cite{Schmitt1998}, similar  processing of OCT images is often discussed to reconstruct displacements produced by additional compression/shearing of the tissue. Subsequent differentiation of the displacement field can be used to determine the elastic-strain field and thus to estimate the shear-modulus distribution \cite{Schmitt1998, Rogowska2004}. 
   
 Despite the apparent simplicity of the idea, the implementation of this approach faced significant difficulties and required highly controllable laboratory conditions in experimental demonstrations (e.g. \cite{Schmitt1998, Rogowska2004}). Indeed, numerical differentiation is rather error-sensitive and therefore requires accurate reconstruction of the displacement field. To fulfill this  requirement, the deformations should not be too small. Otherwise, the pixelized structure of OCT images does not ensure acceptable accuracy even with application of various super-resolution smoothing techniques \cite{Ophir2002,Parker2011}. On the other hand, the tissue strains should not be too large, because distortions of the scatterer patterns in the deformed OCT images reduce the accuracy of the cross-correlation (i.e., produce ``decorrelation noise'' \cite{Ophir2002}). Inevitable other noises in real images introduce additional complications. In what follows we  show that the above-mentioned factors limit the acceptable upper strains to several percentages. This limitation significantly complicates realization of the most practically interesting freehand mode of elastography in OCT. 
   
   Here, we report an alternative variant of mapping the relative stiffness using comparison of OCT images obtained under different degrees of straining.
   We call this variant the correlation-stability (CS) approach, bearing in mind that stiffer regions experience smaller distortions and demonstrate higher cross-correlation (which was also mentioned for ultrasonic scans \cite{Shao2007} ). After presenting numerical simulations and comparison with the conventional DB approach we give some \textit{in vivo} examples of CS maps of the relative elasticity distributions obtained in freehand mode using a spectral-domain OCT scanner. 

For the CS mapping,  the reduction of cross-correlation between the images of the deformed tissue is an informative factor, rather than ``decorrelation noise" reducing the mapping accuracy. There is a similarity with the nonlinear-acoustic approach to detection of cracks (see, e.g., \cite{Zaitsev2006}), in which nonlinear distortions of the sounding field produced by increased nonlinearity of the defects  are intentionally used as a signature of their presence.   
Basically the procedure of cross-correlation between images using a moving (usually rectangular) window   $m_1\times m_2$ in size is similar to that used in refs. \cite{Schmitt1998, Rogowska2004}: 
\begin{equation}
C(n,k)=\frac{\underset{i}\sum\underset{j}\sum(P_{i,j}-\mu_{P})(R_{i+n,j+k}-\mu_{R})}{[\underset{i}\sum\underset{j}\sum(P_{i,j}-\mu_{P})^2\underset{i}\sum\underset{j}\sum(R_{i+n,j+k}-\mu_{R})^2]^{1/2}},
\end{equation}
where $i=1..m_1$,  $j=1..m_2$, the window is moved by $n$ pixels axially and $k$  laterally,  $\mu_{P,R}$  are the mean values in $m_1\times m_2$  areas on images  $P$ and $R$. For ideally coinciding areas $C(0,0)=1$  and tends to zero for uncorrelated ones. For real partially distorted and displaced areas, the position $(n^{*},k^{*})$ is found where the correlation  reaches maximum $C^{*}$ and the field $C^{*}(i,j)$ is plotted as CS-map. 
	
	To give an example of the CS approach let us consider a simulated image of 400x200 pixels in size (typical of OCT images). The ``scatterers'' are formed by putting initially a random value in each pixel.  Then Fourier filtering is applied to smooth the image and obtain inhomogeneities with correlation properties similar to that typical of real OCT images. Namely, the filtering has to ensure the same as in real OCT images level of correlation between non-overlapped correlation windows containing independent scatterers. This background correlation $C^{bg}$ is determined by the number  $q_1\times q_2$ of independent scatterers within the correlation window ($q_{1,2} \leq m_{1,2}$). It can be shown \cite{Matveyev} that 
 $C^{bg}=(\pi/2)^{1/2}/(q_1q_2)^{1/2}$.
In real OCT images, e.g., of a human skin, typically   $C^{bg}\sim0.3-0.2$ for window sizes $m_{1,2} \sim 20-40$ (similar windows were used in \cite{Rogowska2004}). Such values of $C^{bg}$ correspond to sizes of independent scatterers about several pixels, which was ensured for the simulated images. 
	
	The next step is simulation of strained-tissue images. In real experiments the surface of the OCT sensor acts as a rigid piston producing strain fields combining axial and lateral components. Here, for illustration we consider purely shear lateral deformation. This allows one to use a simple analytical solution even for inhomogeneous tissue containing a layer with a contrasting shear modulus $\mu_{2}=b\cdot \mu$, where $\mu=\mu_1=\mu_3$  is the modulus of the surrounding upper and lower layers, and  $b$  is the contrast. If the total sample thickness is  $l=l_1+l_2+l_3$, where $l_2$ is the contrasting-layer thickness (see Fig.\,1a) and the lateral displacement of the tissue 
	surface is $u_0$, then the lateral displacements $u(z)$ inside the sample are 
\begin{equation}
u(z)=\left\{
\begin{array}
{ll}
\frac{u_{0}}{A}\frac{z\mu_{2}}{\mu_{1}l_{2}}, & 0\leq z\leq l_{1}\\
\frac{u_{0}}{A} (  \frac{\mu_{2}l_{1}}{\mu_{1}l_{2}}+\frac{z-l_{1}}{l_{2}
})  , & l_{1}\leq z\leq l_{2} \\
\frac{u_{0}}{A}(  \frac{\mu_{2}l_{1}}{\mu_{1}l_{2}}+1+\frac
{(z-l_{1}-l_{2})\mu_{2}}{\mu_{3}l_{2}} )  , & l_{2}\leq z\leq l_{3}%
\end{array}
\right.
\end{equation}
where $A=1+\frac{\mu_2l_{1}}{\mu_{1}l_{2}}+\frac{\mu_2l_{3}}{\mu_{3}l_{2}}$ and the mean shear strain is $\varepsilon=u_0/l$.
The patterns of the scatterers in the sheared sample can be recalculated using Eqs.\,(2). Continuous subpixel displacements can also be correctly found by using Eqs.\,(2) and applying forward and inverse Fourier transforms of the pattern in combination with the theorem about the Fourier spectrum of shifted functions. 

Figure\,1a shows the initial and deformed patterns of simulated scatterers in which the
\begin{figure}[htb]
\centerline{\includegraphics[width=8.8cm]{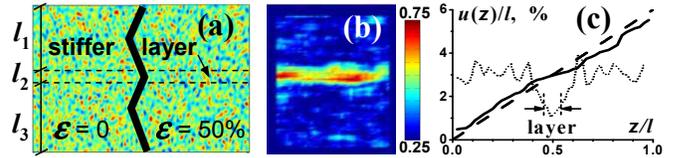}}
\caption{Simulated example of mapping the stiffer layer with contrast $b=3$ and thickness $l_2=20$ pixels. Plot (a) shows the pattern of scatterers with $C^{bg}\approx0.3$ for $\varepsilon=0\%$ and $\varepsilon=50\%$, (b) is the CS image of the stiffer layer obtained by cross-correlating the unstrained pattern with deformed one (see plot (a)). Plot (c) shows the mean displacement $u_{m}(z)$ (dashed line), reconstructed lateral displacement $u(z)$ (solid line) along one of vertical paths, and its derivative averaged over all such paths (dotted line) found by the conventional cross-correlation technique for the mean shear strain $6$\%. The correlation-window size is 20x20 pixels. The images contain no additional noise and the errors are exclusively due to deformation of the scatterer pattern.}
\end{figure}
 contrasting-stiffness layer is not yet visible. An example of CS mapping of this  layer is shown in Fig.\,1b, where the stiffer layer is clearly seen due to increased correlation compared with the surrounding softer layers that experience stronger distortions. Finally, Fig.\,1c shows the reconstructed displacements in a vertical slice using the conventional cross-correlation approach (with parabolic smoothing of correlation peaks to obtain sub-pixel resolution). 
In Fig.\,1c the stiffer layer looks as a butterfly-like deviation of the $u(z)$ function (solid line) from the dashed line $u_{m}(z)$ corresponding to the mean shearing of the entire sample. Conventionally \cite{Schmitt1998}, the so-found displacement field is supposed to be differentiated to obtain the strain field (dotted line shown in Fig.\,1c in arbitrary units),  in which regions with smaller strains correspond to larger stiffness. Such a differentiation is known to introduce additional errors (even if there are no other noises  than the deformation-produced ``decorrelation noise'' \cite{Ophir2002}). In Fig.\,1c  the differentiated dotted line even after averaging over all vertical cuts  demonstrates pronounced irregularities  which are exclusively due to strain-produced distortions of the scatterer pattern. Hereafter, all distortions of the image unrelated to the stress-produced deformation of the scatterer pattern we call “noise”. 

Now we compare the operability regions for the proposed CS mapping and conventional DB elasticity mapping in terms of strain and tolerance to additional noise. In our comparison we even adopt 
the form more favorable for the conventional approach, i.e., avoiding error-sensitive differentiation and bearing in mind that the displacement field itself can already allow one to detect the presence of stiffer regions (which is also mentioned in \cite{Schmitt1998}). Namely, we use the presence of the ``butterfly'' shown in Fig.\,1c as a signature of the presence of the stiffer layer. The mean slope (see the dashed line in Fig.\,1c) can be subtracted for clarity to obtain a horizontally-oriented ``butterfly'' as shown in Fig.\,2a. 
\begin{figure}[htb]
\centerline{\includegraphics[width=8.8cm]{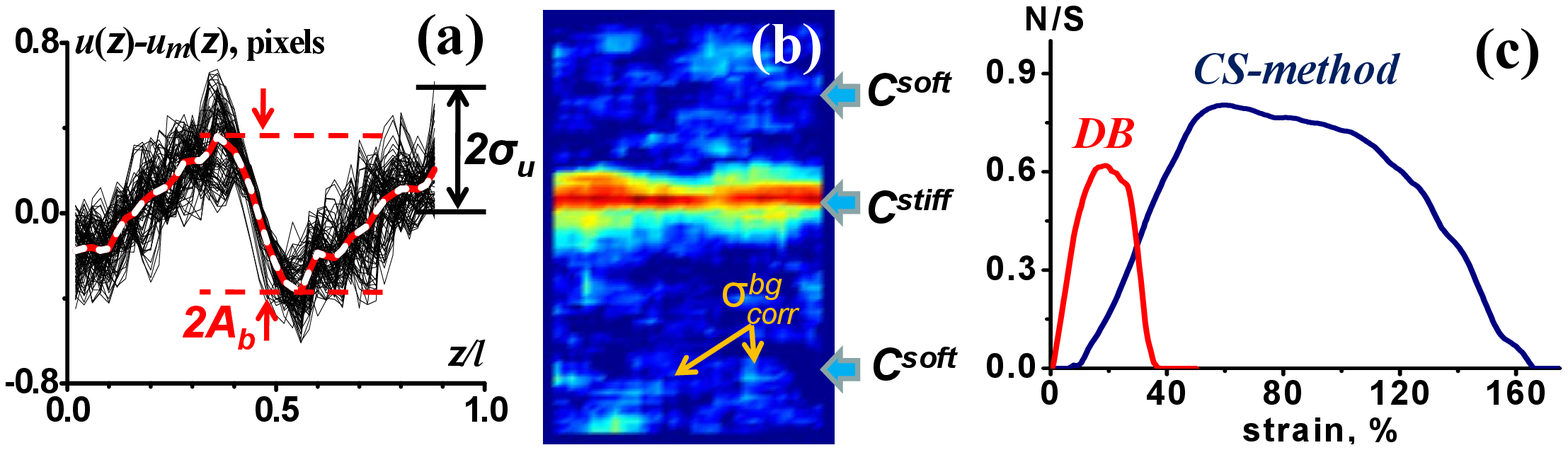}}
\caption{Elucidation to the determining the operability areas for the conventional DB- and CS- mapping. Plot (a) shows a bunch of vertical slices for $u(z)-u_{m}(z)$ with the mean ``butterfly'' of amplitude $A_b$ corresponding to the stiffer layer and irregularities in the reconstructed displacements with variance $\sigma_u$; (b) CS map with regions having difference in the mean correlation $<C^{stiff}-C^{soft}>$ and backgound variance
 $\sigma_{corr}^{bg}$; (c) operability regions of the two methods in terms of strain and noise-to-signal ratio $N/S$ defined as the ratio of square roots of the noise and signal energies (found after subtraction of the respective mean values ). }
\end{figure}
The informative ``butterfly'' with amplitude $A_b$ can be masked by various irregularities characterized by the variance $\sigma_u$ due to the presence of noises and decorrelation distortions of the scatterer pattern in OCT images. As the threshold condition we adopt the often used in statistics ``three-sigma criterion'', i.e.,  $A_b\geq3\sigma_u$. The variance 
$\sigma_u$ of the irregularities is estimated for each given strain by gradually increasing the amplitude of the generated additional noise and then comparing $\sigma_u$ with $A_b$.
For the CS mapping, we make a similar comparison between the cross-correlation level $C^{stiff}$ for the stiffer layer and $C^{soft}$ for the surrounding softer areas. These two levels of cross-correlation are characterized by their respective average values  and the background cross-correlation variance  $\sigma_{corr}^{bg}$ (which is due to irregularities in the cross-correlation coefficient that are always present even in the absence of the stiffer layer). For a given strain, we vary the noise level and again determine the threshold difference $<C^{stiff}-C^{soft}>=3\sigma_{corr}^{bg}$. The so-found curves delimiting the operability regions for the DB and CS mapping are shown in Fig.\,2c. In agreement with other authors \cite{Ophir2002, Kirkpatrick2006} we found that the conventional DB mapping works for fairly small (below $\sim10\%$) strains, whereas the CS mapping works in a much wider range of larger strains ($\sim20-150\%$), which is favorable for freehand operation.  Besides, CS mapping is noticeably more tolerant to noises.   
\begin{figure}[htb]
\centerline{\includegraphics[width=7.9cm]{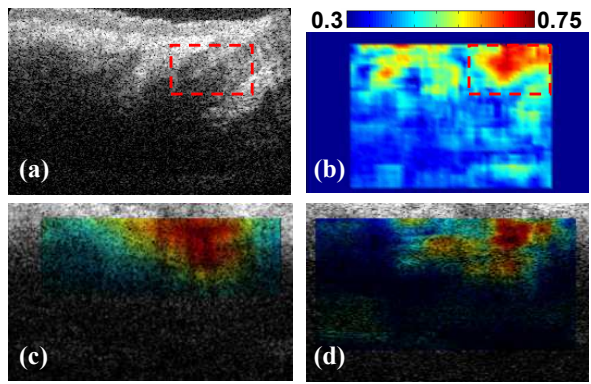}}
\caption{Two \textit{in vivo} examples of CS maps 
where red color shows the stiffer regions. Panel (a) shows a conventional OCT image  where a hair root is hardly seen and (b) is the corresponding CS map for correlation-window size $20\times20$ pixels. Panels (c) and (d) show OCT images of another bulb superposed with the elastographic CS maps with different resolution obtained using windows $40\times40$ and $20\times20$ pixels, respectively.}
\end{figure}
For the demonstrations, we used a recently developed OCT spectral-domain scanner based on techniques described in \cite{Gelikonov2009a,Gelikonov2009b}. It acquires $480\times289$\,pixel images with a rate of $21$\,fps, axial resolution $10$\,$\mu$m and lateral one $20$\,$\mu$m. The fiber-optic OCT probe is 2 mm in diameter and acts as a piston producing deformation of the tissue during freehand operation. The images are fed in a PC in real time and after filtering, amplitude normalization, and averaging of a few frames are cross-correlated with a reference one. For \textit{in vivo} examples we have chosen human cheek skin where hair roots serve as stiffer inclusions. The resulting CS maps are shown in Fig.\,3, where the stiffer regions are clearly seen. For these examples, the average strain created in the tissue by pressing the probe is about $20-50$\% in agreement with the above-found operability region for CS mapping.

Thus the
performed analysis and tests 
demonstrate that the rather straightforward CS approach can readily be implemented in sufficiently high-speed OCT scanners in freehand mode to obtain relative-stiffness maps similar to those used in medical ultrasonics \cite{Parker2011}.

Support of the grant 11.G34.31.0066 of the Russian Federation Government is acknowledged.

\newpage
\section*{Informational Fourth Page}

\textbf{Full-form references}

1. J. Schmitt, "OCT elastography: imaging microscopic deformation and strain of tissue.," Optics express \textbf{3}, 199–211 (1998).

2. J. Rogowska, N. A. Patel, J. G. Fujimoto, and M. E. Brezinski, "Optical coherence tomographic elastography technique for measuring deformation and strain of atherosclerotic tissues," Heart \textbf{90}, 556–562 (2004).


3. S. J. Kirkpatrick, R. K. Wang, and D. D. Duncan, "OCT-based elastography for large and small deformations.," Optics express \textbf{14}, 11585–97 (2006).

4. E. Bossy, A. R. Funke, K. Daoudi, A.-C. Boccara, M. Tanter, and M. Fink, "Transient optoelastography in optically diffusive media," Applied Physics Letters \textbf{90}, 174111 (2007).

5. S. G. Adie, X. Liang, B. F. Kennedy, R. John, D. D. Sampson, and S. A. Boppart, "Spectroscopic optical coherence elastography.," Optics express \textbf{18}, 25519–34 (2010).

6. B. F. Kennedy, X. Liang, S. G. Adie, D. K. Gerstmann, B. C. Quirk, S. A. Boppart, and D. D. Sampson, "In vivo three-dimensional optical coherence elastography," Optics Express \textbf{19}, 6623–6634 (2011).

7. C. Sun, B. Standish, and V. X. D. Yang, "Optical coherence elastography: current status and future applications.," Journal of Biomedical Optics \textbf{16}, 043001 (2011).

8.R. K. Wang, Z. H. Ma, and S. J. Kirkpatrick, "Tissue Doppler optical coherence elastography for real time strain rate and strain mapping of soft tissue,”  Appl. Phys. Lett. \textbf{89}, 144103 (2006).

9.  B. F. Kennedy, A. Curatolo, T. R. Hillman, C. M. Saunders, and  D. D. Sampson, 
“Speckle reduction in optical coherence tomography images using tissue viscoelasticity”, 
 J. Biomed. Opt. \textbf{16}, 020506  (2011).

10. K. J. Parker, M. M. Doyley, and D. J. Rubens, "Imaging the elastic properties of tissue: the 20 year perspective.," Physics in Medicine and Biology \textbf{56}, R1–R29 (2011).

11. J. Ophir, S. Alam, B. Garra, and F. Kallel, "Elastography: imaging the elastic properties of soft tissues with ultrasound," J. Medical Ultrasonics \textbf{29}, 155–171 (2002).

12. J. Shao, J. Bai, L. Cui, J. Wang, Y. Fu, K. Liu, and S. Feng, "Elastographic Evaluation of the Temporal Formation of Ethanol-Induced Hepatic Lesions," J. Ultrasound Med. \textbf{26}, 1191–1199 (2007).

13. V. Y. Zaitsev, V. E. Nazarov, and V. I. Talanov, "“Nonclassical” manifestations of microstructure-induced nonlinearities: new prospects for acoustic diagnostics," Physics Uspekhi \textbf{49}, 89--94 (2006).
 
14. J.S.Bendat, A.G.Piersol. “Random data: Analysis and 
measurement procedures”, J.Wiley\&Sons, NY(1986).

15. V.M. Gelikonov, G.V. Gelikonov, P.A. Shilyagin, ``Linear-Wavenumber Spectrometer for High-Speed Spectral-Domain Optical Coherence Tomography'', Optics and Spectroscopy  \textbf{106}, N. 3, 459--465 (2009).

16. V.M. Gelikonov, G.V. Gelikonov, I.V. Kasatkina, D.A. Terpelov, P.A. Shilyagin, ``Coherent Noise Compensation in Spectral-Domain Optical Coherence Tomography'',  Optics and Spectroscopy,  \textbf{106}, N. 6,  895--900 (2009).

\end{document}